\documentclass[prl,twocolumn,showpacs,amsmath,amssymb, superscriptaddress]{revtex4-1}
\usepackage{graphicx,braket}
\usepackage{color}
\begin{document}

\title{Massless Dirac Equation from Fibonacci Discrete-Time Quantum Walk}

\author{Giuseppe Di Molfetta}
\email{giuseppe.dimolfetta@ens.fr}
\altaffiliation{Contributed equally to this work.}
 \affiliation{Research Center of Integrative Molecular Systems (CIMoS), Institute for Molecular Science, 38 Nishigo-Naka, Myodaiji, Okazaki, Aichi 444-8585, Japan.}
 \affiliation{LERMA, UMR 8112, UPMC - Paris 6 and Observatoire de Paris, 61 Avenue de l'Observatoire, 75014 Paris, France.}
\author{Lauchlan Honter}
 \altaffiliation{Contributed equally to this work.}
 \affiliation{Research Center of Integrative Molecular Systems (CIMoS), Institute for Molecular Science, 38 Nishigo-Naka, Myodaiji, Okazaki, Aichi 444-8585, Japan.}
 \affiliation{School of Physics, The University of Western Australia, 35 Stirling Hwy, Crawley, Perth WA 6009, Australia.}
\author{Ben B. Luo}
\altaffiliation{Contributed equally to this work.}
 \affiliation{Research Center of Integrative Molecular Systems (CIMoS), Institute for Molecular Science, 38 Nishigo-Naka, Myodaiji, Okazaki, Aichi 444-8585, Japan.}
 \affiliation{School of Physics, The University of Western Australia, 35 Stirling Hwy, Crawley, Perth WA 6009, Australia.}
\author{Tatsuaki Wada}
 \affiliation{Department of Electrical and Electronic Engineering, Ibaraki University, Hitachi, Ibaraki 316-8511, Japan.}
\author{Yutaka Shikano}
 \email{yshikano@ims.ac.jp}
 \affiliation{Research Center of Integrative Molecular Systems (CIMoS), Institute for Molecular Science, 38 Nishigo-Naka, Myodaiji, Okazaki, Aichi 444-8585, Japan.}
 \affiliation{Institute for Quantum Studies, Chapman University, 1 University Dr., Orange, California 92866, USA.}
\date{\today}

\begin{abstract}
	Discrete-time quantum walks can be regarded as quantum dynamical simulators since they can simulate spatially discretized Schr\"{o}dinger, massive 
	Dirac, and Klein-Gordon equations. Here, two different types of Fibonacci discrete-time quantum walks are studied analytically. The first is the Fibonacci coin sequence 
	with a generalized Hadamard coin and demonstrates six-step periodic dynamics. The other model is assumed to have three- or six-step periodic dynamics with the Fibonacci sequence. 
	We analytically show that these models have ballistic transportation properties and continuous limits identical to those of the massless Dirac 
	equation with coin basis change. 
\end{abstract}
%05.45.Mt   Quantum chaos; semiclassical methods
%03.65.-w   Quantum mechanics
%02.30.Jr    Partial differential equations
\maketitle

%%%%%%%%%%%%%%%%%%%%%%  Introduction   %%%%%%%%%%%%%%%%%%%%%%%%%%%%%%%%
Discrete-time quantum walks (DTQWs) are defined as quantum-mechanical analogues of classical random walks. The concept of DTQWs was first considered by Feynman \cite{Feynman1965} and 
then introduced in greater generality by Refs. ~\cite{Aharonov1993, Meyer1996, Gudder1988}. They have been realized experimentally in Refs. \cite{Cardano2014, Broome2010, Kitagawa2012, Schreiber2012, 
Schmitz2009, Zahringer2010, Schreiber2010, Karski2009, Sansoni2012, Sanders2003, Perets2008, Crespi2013, Jeong2013, Fukuhara2013, Xue2014, Manouchehri2014} and are important in many fields, from 
fundamental quantum physics \cite{Perets2008, DiMolfetta2014, Shikano2010, Chandrashekar2008} to quantum algorithm~\cite{Ambainis2007, Magniez2006} and condensed matter 
physics~\cite{Bose2003, Aslangul2005, Oka2005, Bose2007, Kitagawa2010}. Previously, it has been shown that several DTQWs on a line admit a continuous limit identical to 
the propagation equations of a massive Dirac fermion~\cite{Strauch2007, Bracken2007, Sato2010, Chandrashekar2010, DiMolfetta2011} and those of massless Dirac fermion equations~\cite{Bracken2007, DiMolfetta2014}. 
Furthermore, the relationship between DTQWs and artificial electric and gravitational fields has been shown~\cite{DiMolfetta2013, DiMolfetta2014}. Thus, DTQWs can be regarded as quantum 
dynamical simulators \cite{Shikano2013}.
Additionally, it is well known that the classical random walk leads to a diffusive behavior characterized by the time evolution of the standard deviation, with $\sigma(t) \sim t^{1/2}$, 
while the standard DTQW leads to ballistic behavior, as $\sigma(t) \sim t$. Further, the standard DTQW can be considerably enriched by generalizing the
quantum coin operator and arranging it along different sequences. It has already been shown that quasi-periodic coin sequences induced by the Fibonacci sequence 
lead to sub-ballistic behavior, whereas random sequences lead to diffusive spreading~\cite{Ribeiro2004}. 
Here, we consider two different Fibonacci DTQWs with periodic coin sequences. The first model (FDTQW-I) considers a time-dependent quantum coin following 
the Fibonacci sequence, while the second model (FDTQW-II) considers a modified version of the unitary operator first defined in Ref. \cite{Ribeiro2004}, where the 
Fibonacci sequence is applied to the step operator. We show numerically and analytically that the continuous limit of these models reduces to a massless Dirac 
equation in $(1+1)$ dimensions. 

%%%%%%%%%%%%%%% Model Explanation   %%%%%%%%%%%%%%%%%%%%%%%%%%%%%%%%%%%%
Let us consider the two dimensional spin state $\Psi_{m,j}\in\mathbb{C}^{2}$, spanned by the orthonormal basis ($b_{u}, b_{d}$), and defined by its discrete one dimensional 
position $m\in\mathbb{Z}$  and discrete time $j \in \mathbb{N}_0$. The  standard DTQW's time evolution is given by the application of the quantum coin 
operator (QCO) $\hat{C}$ on $\Psi_{m,j} = u_{m,j} b_u + d_{m,j} b_d = \left( \begin{array}{c} u_{m,j} \\ d_{m,j} \end{array} \right)$,
followed by the chiral-dependent translation operator $\hat{T}$, which is defined as
\begin{equation}
\left(
\begin{array}{cc}
u_{m-1,j} \\
d_{m+1,j}\\
\end{array}\right) = \hat{T}\left(
\begin{array}{cc}
u_{m,j} \\
d_{m,j}  \\
\end{array}\right).
\end{equation}
Here, we introduce the simplest quantum coin, the generalized Hadamard coin, which is expressed as
\begin{equation}
\hat{C} (\theta)=
\left(
\begin{array}{cc}
\cos (\theta ) & \sin (\theta ) \\
\sin (\theta ) & -\cos (\theta ) \\
\end{array}
\right),
\end{equation}
where  $\theta\in[0,2 \pi]$.
The one-step discrete time evolution is then given by
\begin{equation}
\begin{pmatrix} u_{m,j+1}\\ d_{m,j+1}\end{pmatrix} = \hat{T}\hat{C}(\theta)\begin{pmatrix} u_{m,j}\\ d_{m,j}\end{pmatrix}\ . \label{generalform}
\end{equation}

First, we consider FDTQW-I, which is the simplest case as only the QCO is defined as a Fibonacci series. Here, 
\begin{equation}
	\begin{aligned}
		\hat{U}_{j}=\hat{T}\hat{C}_{j}\ , \qquad
	j \in \mathbb{N}_{0}\ ,\qquad
\hat{C}_{j+1} = \hat{C}_{j} \hat{C}_{j-1}\ ,
	\end{aligned}
\end{equation}
with the initial conditions
\begin{equation}
	\begin{aligned}
		\hat{C}_{0}=\hat{C}(\alpha)\ ,\qquad
		\hat{C}_{1}=\hat{C}(\alpha)\hat{C}(\beta),
	\end{aligned}
\end{equation}
where $j$ is the time step.
On considering the DTQW acted upon by the Fibonacci coin series, 
we analytically find that the time evolution of the coin is cyclic with period $6$. 
These coin operators then reduce to
\begin{equation}
\begin{aligned}
\\
\hat{C}_{0}&=\begin{pmatrix} \cos(\alpha) & \sin(\alpha) \\ \sin(\alpha) & -\cos(\alpha) \end{pmatrix}\ ,\\
\hat{C}_{1}&=\begin{pmatrix} \cos(\alpha-\beta) & -\sin(\alpha-\beta) \\ \sin(\alpha-\beta) & \cos(\alpha-\beta) \end{pmatrix}\ ,\\
\hat{C}_{2}&=\begin{pmatrix} \cos(2\alpha-\beta) & \sin(2\alpha-\beta) \\ \sin(2\alpha-\beta) & -\cos(2\alpha-\beta) \end{pmatrix}\ ,\\
\hat{C}_{3}&=\begin{pmatrix} \cos(\alpha) & \sin(\alpha) \\ \sin(\alpha) & -\cos(\alpha) \end{pmatrix}\ ,\\
\hat{C}_{4}&=\begin{pmatrix} \cos(\alpha-\beta) & \sin(\alpha-\beta) \\ -\sin(\alpha-\beta) & \cos(\alpha-\beta) \end{pmatrix}\ ,\\
\hat{C}_{5}&=\begin{pmatrix} \cos(\beta) & \sin(\beta) \\ \sin(\beta) & -\cos(\beta) \end{pmatrix}.\\
\end{aligned}
\end{equation}
Here, the collection $W^{n}_{j}=(\Psi_{m,k})_{m\in\mathbb{Z}, k=nj}$ is defined for $j \in \mathbb{N}$ and $n \in \{ 0, 1, 2, 3, 4, 5 \}$, and represents the state of the walk at time $k=nj$. For any given $n$ 
we define $S^{n}=(W^{n}_{j})_{j\in\mathbb{Z}}$, where $S^{n}$ represents the entire history of the walk observed through a stroboscope of period $n$. Successive application of the $6$ unitary operators to an initial state 
then gives the stroboscopic recursion equations for $S^{6}$. The discrete-step equations for $S^6$ read
\begin{flalign}
u_{m,j+6} &= \sum\limits_{k=-3}^3( A_{2k}(\alpha,\beta) u_{m+2k,j} \nonumber \\ 
& \ \ \ \ \ \ \ \ \ \ \ \ \ \ \ \ \ \ + B_{2k}(\alpha,\beta) d_{m+2k,j}), \label{rreq1} \\
d_{m,j+6} &= \sum\limits_{k=-3}^3( B_{-2k}(-\alpha,-\beta) u_{m+2k,j} \nonumber \\ 
& \ \ \ \ \ \ \ \ \ \ \ \ \ \ \ \ \ \ + A_{-2k}(-\alpha,-\beta) d_{m+2k,j}). \label{rreq2}
\end{flalign}
Here, the index $k$ $\in$ $\{ -6, -4, -2, 0, 2, 4, 6 \}$ and 
the coefficients $A_k$, $B_k$ $\in$ $\mathbb{R}$ are explicitly given as 
\begin{small}
\begin{flalign}
A_{-6}&=  c^2_\alpha c_\beta c^2_{\alpha -\beta} c_{2\alpha-\beta} \\ \nonumber
A_{-4}&= -\frac{1}{4} c_\alpha c^2_{\alpha -\beta} (c_{\alpha -2 \beta}+3 c_{3 \alpha -2 \beta} - 5 c_{\alpha} + c_{3 \alpha}) \\ \nonumber
A_{-2}&= \frac{1}{16} (-6 c_{2 (\alpha -\beta )} + 4 c_{4 (\alpha -\beta )} - c_{2 (\alpha +\beta )}- c_{2 (\alpha -2 \beta )} \\  \nonumber
&+ 2 c_{4 \alpha -2 \beta } - c_{6 \alpha -2 \beta } + c_{6 \alpha -4 \beta} - 2 c_{4 \alpha} - 2 c_{2\beta }+6)\\ \nonumber
A_{\ 0}&= \frac{1}{4} c^2_{\alpha -\beta} (-6 c_{2 (\alpha -\beta )}+c_{4 \alpha -2 \beta }-2 c_{2\alpha }+c_{4 \alpha }+ c_{2 \beta}+5)\\ \nonumber
A_{\ 2}&= \frac{1}{8} c^2_{\alpha -\beta} (6 c_{2 (\alpha -\beta )}-3 c_{4 \alpha -2 \beta }-2 c_{2 \alpha }+c_{4 \alpha}+c_{2 \beta}-3)\\ \nonumber
A_{\ 4}& = \frac{1}{2} s_{2 \alpha} s_{\beta} c^2_{\alpha-\beta} c_{2 \alpha -\beta}\\ \nonumber
A_{\ 6}& = 0 \\
B_{-6}&= \frac{1}{2} s_{2 \alpha} c_{\beta} c^2_{\alpha -\beta} c_{2 \alpha -\beta} \\ \nonumber
B_{-4}&= \frac{1}{8} (s_{2 \alpha} s_{\beta} (s_\beta- s_{4 \alpha -3 \beta})+c_\beta (3 s_{2 \alpha -\beta}-s_{4 \alpha -\beta}+\\ \nonumber
&3 s_{4 \alpha -3 \beta}-s_{6 \alpha -3 \beta})) \\ \nonumber
B_{-2}&= \frac{1}{8}((c_{2 \alpha}-3) s_{4 \alpha -4 \beta}-2 s_{4 \alpha}c^2_{\alpha -\beta})\\ \nonumber
B_{\ 0}&= \frac{1}{16} (-s_{2 \alpha -4 \beta }+4 s_{4 \alpha -4 \beta}+s_{6 \alpha -4 \beta}+2 c_\alpha (s_{\alpha +2 \beta}-\\ \nonumber 
&s_{\alpha -2 \beta}-3 s_{3 \alpha -2 \beta}+ s_{5 \alpha -2 \beta})-2 s_{2 \alpha}+2 s_{4 \alpha})\\ \nonumber
B_{\ 2}&=-\frac{1}{16} c_{\alpha} (s_{3 \alpha -4 \beta}+3 s_{5 \alpha -4 \beta}+8 s^3_{\alpha} c_{2 \alpha -2 \beta} +4 s_{\alpha}-2 s_{3 \alpha} ) \nonumber \\ \nonumber
B_{\ 4}& = -c^2_\alpha s_\beta c^2_{\alpha -\beta} c_{2 \alpha -\beta} \\ \nonumber
B_{\ 6}& = 0
\end{flalign}
\end{small}
with $c_\theta := \cos (\theta)$ and $s_\theta := \sin (\theta)$.\\

\begin{figure}[h!]
\centering
\includegraphics[width=1 \columnwidth]{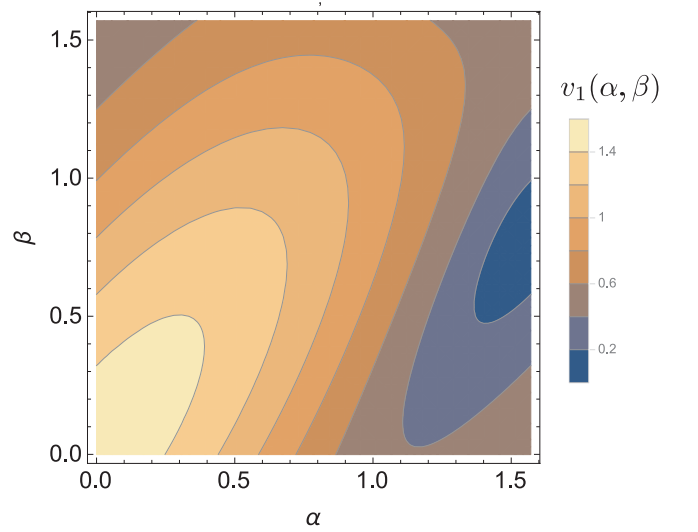}
\caption{Velocity contour plot for the Fibonacci sequence on the QCO, with $\alpha$ and $\beta$ $\in [0, \frac{\pi}{2}]$.}
\label{fig:vel1}
\end{figure}

Let us define the time and space variables, $t_j =j \Delta t$ and $x_m = m\Delta x$, where $\Delta t$ and $\Delta x$ are the time and space steps, respectively.
As $\Delta t$ and $\Delta x$ tend to zero, this allows us to take a Taylor expansion of the recursion relations for the DTQW and, hence, derive a pair of 
partial differential equations (PDEs). To take the continuous limit, we define
\begin{equation}
\begin{aligned}
\Delta t &= \epsilon\ ,\\
\Delta x &= \epsilon^{\gamma}\ ,\\
\end{aligned}
\end{equation}
where $\epsilon$ is an infinitesimal and $\gamma>0$ is a scaling parameter. The difference between the two expressions is to account for the fact that $\Delta t$ and $\Delta x$ may tend to $0$ differently.
Then, taking a Taylor expansion about $\epsilon$ up to the leading orders of Eqs (\ref{rreq1}) and (\ref{rreq2}), we obtain the following  
\begin{multline}
u(x,t) + 6 \epsilon\partial_{t}u(x,t) = u(x,t)+\\
+ \epsilon^\gamma  (p_1\partial_{x}u(x,t)+ p_2 \partial_{x}d(x,t))+O(\epsilon^{2}),
\end{multline}
\begin{multline}
d(x,t)+ 6 \epsilon \partial_{t}d(x,t) = d(x,t)+\\
+ \epsilon^\gamma (p_2 \partial_{x}u(x,t) - p_1 \partial_{x}d(x,t)) + O(\epsilon^{2}),
\end{multline}
where
\begin{small}
\begin{align}
p_1&=   \sum\limits_{k=-3}^3 \frac{2k}{6} A_{2k}(\alpha,\beta) = -\frac{1}{6}(c_{4\alpha-2\beta}+2c_{2\alpha}+c_{2\beta}+2),\\
p_2&=   \sum\limits_{k=-3}^3 \frac{2k}{6} B_{2k}(\alpha,\beta)= -\frac{2}{3}(s_{2\alpha}c^2_{\alpha-\beta}).
\end{align}
\end{small}
Choosing scaling of $\gamma=1$ and then taking the limit as $\epsilon\rightarrow0$, we obtain the following pair of PDEs:
\begin{equation}
\begin{aligned}
\partial_{t}u(x,t) &= p_{1} \partial_{x}d(x,t) +p_{2} \partial_{x}u(x,t)\ ,\\
\partial_{t}d(x,t) &= p_{2} \partial_{x}d(x,t)
-p_{1}\partial_{x}u(x,t) .\\
\end{aligned}
\end{equation}
This set of equations can be then be recast, such that
\begin{equation}
\begin{aligned}
\mathbb{I}\partial_t \Psi + &P \partial_x \Psi = 0\ , & &
P=\begin{pmatrix} p_{1} & p_{2} \\ 	p_{2} & -	p_{1} \end{pmatrix}\ ,\\
\end{aligned}
\label{fineq}
\end{equation}
where $\mathbb{I}$ is the $2\times 2$ identity matrix. 
To diagonalize the operator acting on $\Psi$, we perform a change of basis from ($b_{u}$,$b_{d}$) to the new 
basis, ($\overline{b_{u}},\overline{b_{d}}$), with  $\overline{\Psi}=\overline{u}\overline{b}_u+\overline{d}\overline{b}_{d}$. 
The new basis components are
\begin{align}
\overline{b_{u}} &= \frac{1}{Z} \left( \frac{p_2}{\omega - p_1} b_{u} + b_{d} \right), \\
\overline{b_{d}} &= \frac{1}{Z} \left( -\frac{p_2}{\omega + p_1} b_{u} + b_{d} \right),
\end{align}
with $\omega = \sqrt{p_1^2+p_2^2}$ and $Z$ a normalized constant.
Hence, equation (\ref{fineq}) in the new basis reads
\begin{equation}
\mathbb{I} \partial_t \overline{\Psi} + v(\alpha,\beta)\sigma_{z} \partial_x \overline{\Psi} =0, \label{modeq}
\end{equation}
where
\begin{small}
\begin{align}
v_1(\alpha,\beta) &= \frac{\sqrt{8c^2_{\alpha}c_{2\alpha-2\beta}+c_{4\alpha-4\beta}+4c_{2\alpha}+5}}{3\sqrt{2}},
\label{eq:v1}
\end{align}
\end{small} can be seen as the propagation velocity of the continuous limit distribution and $\sigma_z$ is the third Pauli matrix. 
We may also evaluate the standard deviation $\sigma_j$ of the probability distribution depicted in Fig. \ref{fig:vel1} as a function of time, by 
considering the exponent $\eta(\alpha,\beta)$ in $\sigma_j \sim j^{\eta(\alpha,\beta)}$. For the FDTQW-I case, we observe ballistic behavior 
for general $(\alpha,\beta)$, that is, $\eta(\alpha,\beta)=1$.

\begin{figure}[h!]
\centering
\includegraphics[width=1 \columnwidth]{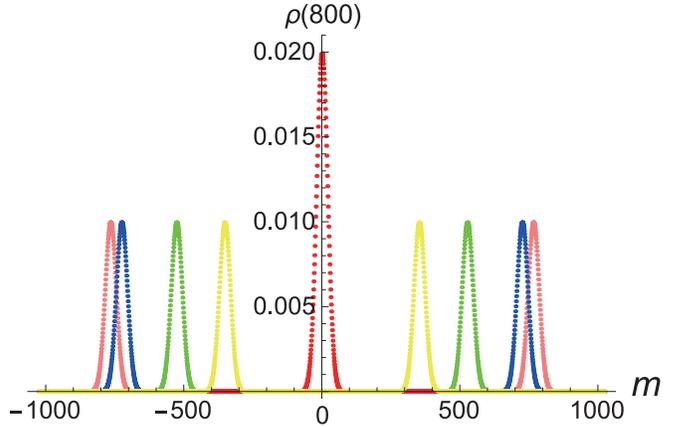}
\caption{Density profile of FDTQW-I at time step, $j = 800$, $\rho (j) = \mid\Psi_{m,j} \mid^2$, versus space step, $m$. The resolution (i.e. $\epsilon$ value) is $n=2^{11}$. The dashed red line indicates $\alpha$ = $\pi$/2 and $\beta$ = $\pi/4$; the dashed yellow line is for $\alpha$ = $\pi$/3 and $\beta$ = $\pi/6$; the dashed green line represents $\alpha$ = $\pi/4$ and $\beta$ = $\pi/8$; the dashed blue line indicates $\alpha$ = $\pi$/8 and $\beta$ = $\pi/16$; and the dashed pink line represents $\alpha$ = $\pi$/12 and $\beta$ = $\pi/24$. The initial condition is $\Psi(0,x) = \sqrt{N_0(x)}(b_u + I b_d)$, where $N_0(x)$ is a Gaussian function of width $20 \Delta x$ and $\Delta x$ = 2 $\pi/n$.}
\label{fig:prob1}
\end{figure}

Now consider Eq. (\ref{modeq}) in the covariant form, which is expressed as 
\begin{equation}
i (\gamma^0 \partial_0 + \gamma^1 \partial_{1}) \overline{\Psi}=0,
\end{equation}
where $\gamma^0 = \sigma_x$ and $\gamma^1 = - i \sigma_y$ are the usual gamma matrices, $\partial_0=\partial_s, \ \partial_1=\partial_{\tilde{x}}$, and the rescaled coordinate $\tilde{x}$ = $x / v_1(\alpha, \beta)$. This equation can now be interpreted as the massless Dirac equation in the 1+1 space-time dimension. In Fig. \ref{fig:prob1}, we observe the density profile of the FDTQW-I case at a time step of $j=800$, with a symmetric Gaussian 
initial condition that is sufficiently regular and large with respect to the lattice interval $\Delta x$. A truly ballistic propagation can be noted, as the continuous limit suggests. This result confirms that FDTQW-I can 
be used to simulate massless Dirac dynamics.

The FDTQW-II previously defined in Refs. \cite{Ribeiro2004,Romanelli2009} can be expressed in the most general case as a unitary evolution, and the 
unitary evolution operator can then be defined using the following Fibonacci sequence
\begin{equation}
\hat{U}_{j}=\hat{U}_{j-1}\hat{U}_{j-2}\ , \label{fdtqw2}
\end{equation}
with the initial conditions 
\begin{equation}
\hat{U}_{0}=\hat{T}\hat{C}(\alpha),\qquad
\hat{U}_{1}=\hat{T}\hat{C}(\alpha)\hat{T}\hat{C}(\beta).
\end{equation}
Since each step operator contains increasing numbers of translation operators, the boundary size increases at an exponential rate. 
To account for this, we now parametrize time using the number of translation operators that have thus far been applied. It can then be noted that 
this DTQW has a quasi-periodic coin sequence. Thus, we define a new parameter for time $r$, such that $s_r=r \Delta t$, where $r=\sum_{n=0}^{j-1}F(n)$ and 
$F(n)$ is the Fibonacci sequence with initial conditions, $F(0)=1$ and $F(1)=2$. This means how many translation operators a quantum walker is operated.
This allows us to easily study the continuous limits of Eq (\ref{generalform}) for the operator expressed in Eq (\ref{fdtqw2}).
For the continuous limit to exist, we first require that $\Psi(s_r+\tau\Delta t)\rightarrow\Psi(s_r)$. It can be shown that this 
is true only when $\tau \in 6\mathbb{N}_{0}$, that is, when $\tau$ is any positive integer multiple of $6$, as the sequence of 
unitary operators then reduces to the identity operator. We assume the simplest periodic case, for which $\tau = 6$, and 
the discrete-time equations read
\begin{flalign}
u_{m,r+6} & = \sum\limits_{k=-3}^3( A_{2k}(\alpha,\beta) u_{m+2k,r} \nonumber \\ 
& \ \ \ \ \ \ \ \ \ \ \ \ \ \ \ \ \ \ + B_{2k}(\alpha,\beta) d_{m+2k,r}), \\
d_{m,r+6} & = \sum\limits_{k=-3}^3 (B_{-2k}(-\alpha,-\beta) u_{m+2k,r} \nonumber \\ 
& \ \ \ \ \ \ \ \ \ \ \ \ \ \ \ \ \ \ + A_{-2k}(-\alpha,-\beta)  d_{m+2k,r}),
\end{flalign}
where $A_k \in \mathbb{R}$ and $B_k \in \mathbb{R}$ are
\begin{flalign}
A_{-6}&= c^4_\alpha c^2_\beta \\ \nonumber
A_{-4}&= c^2_\alpha s_\alpha (c^2_\beta s_\alpha + 2 c_\alpha s_{2 \beta}) \\ \nonumber
A_{-2}&= -\frac{1}{8}s_{2\alpha}(-2 s_{2 \alpha} + s_{2(\alpha-\beta)}+5 s_{2(\alpha+\beta)}) \\ \nonumber
A_{\ 0}& = -\frac{1}{8}(3 + c_{4\alpha} - (1+3c_{4\alpha})c_{2\beta}-16c_{\alpha}s_{\alpha}^3s_{2\beta}) \\ \nonumber
A_{\ 2}&= c^2_\beta s^4_{\alpha} - 2 c^3_{\alpha}c_{\beta}s_{\alpha}s_{\beta} + c_\alpha s_\alpha^3 s_{2\beta} \\ \nonumber
A_{\ 4}&=  c^2_{\alpha}c^2_{\beta}s^2_{\alpha}\\ \nonumber
A_{\ 6}&= 0 \\ 
B_{-6}&=  c^3_\alpha c^2_{\beta} s_\alpha \\ \nonumber
B_{-4}&= c_\alpha c_\beta (s^3_\alpha c_\beta + c_\alpha (1-2 c_{2\alpha}) s_\beta)\\ \nonumber
B_{-2}&= \frac{1}{8} (s_{4\alpha} (3 c_{2 \beta}-1)-4 s^2_\alpha (2 c_{2 \alpha}+1) s_{2\beta})\\ \nonumber
B_{\ 0}&=  \frac{1}{8} (4 s^2_\alpha (2 c_{2\alpha}+1) s_{2\beta} + s_{4 \alpha} (1-3 c_{2\beta}))\\ \nonumber
B_{\ 2}&=  c_\alpha c_\beta (c_\alpha (2 c_{2 \alpha}-1) s_{\beta}-s^3_\alpha c_\beta)\\ \nonumber
B_{\ 4}&= - s_\alpha c^3_\alpha c^2_\beta\\ \nonumber
B_{\ 6}&= 0.\nonumber 
\end{flalign}

\begin{figure}[h!]
\centering
\includegraphics[width=1 \columnwidth]{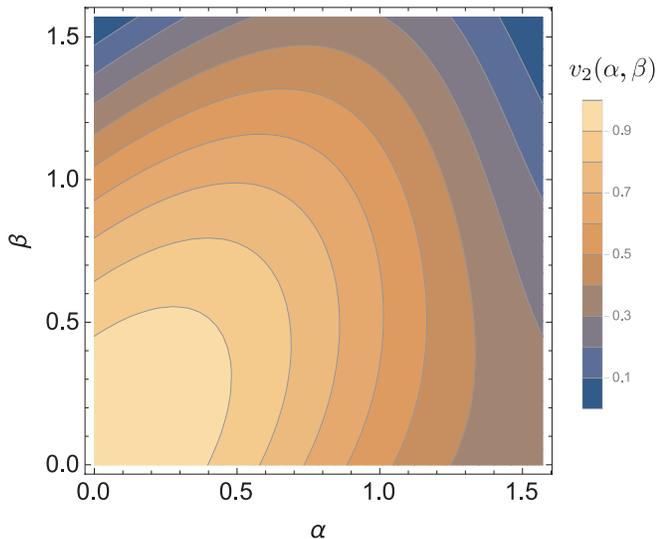}
\caption{Contour plot for velocity in the case in which Fibonacci sequence is on the step operator for the three periodic Fibonacci sequence.}
\label{fig:vel2}
\end{figure}

Taking the continuous limit of $S^6$ about $\epsilon$ (as in the previous section) and noting that the zeroth order terms 
cancel, we arrive at the first order term expressions for this system  
\begin{equation}
\begin{aligned}
\partial_{s}u(x,s) &=p_{1} \partial_{x}d(x,s)+p_{2} \partial_{x}u(x,s), \\
\partial_{s}d(x,s) &= p_{2} \partial_{x}d(x,s)-p_{1} \partial_{x}u(x,s),
\end{aligned}
\end{equation}
and 
\begin{align}
p^\prime_1&= \sum\limits_{k=-3}^3 \frac{2k}{6}  A_{2k}(\alpha,\beta) = 
- \frac{1}{3}(c^2_{2\alpha-\beta}+2 c_{\beta}c_{2\alpha-\beta}) \\
p^\prime_2&= \sum\limits_{k=-3}^3 \frac{2k}{6}  B_{2k}(\alpha,\beta)= \frac{1}{3}s_{2\alpha-\beta}(c_{2\alpha-\beta}+2c_{\beta}).
\end{align}
As in the previous case, we can rewrite these equations in matrix form, such that 
\begin{equation}
\begin{aligned}
&\partial_s \Psi + P \partial_x \Psi =0 , \\
P = &\left(
\begin{array}{cc}
p^\prime_{1} & p^\prime_{2} \\
p^\prime_{2}& -p^\prime_{1}\\
\end{array}
\right)\ ,\\
\end{aligned}
\label{fineq2}
\end{equation}
Again, we diagonalize the differential operator acting on $\Psi$, which spans $\Psi$ itself on the new basis, ($\overline{b^\prime_{u}},\overline{b^\prime_{d}}$). 
The new basis components read
\begin{align}
\overline{b^\prime_{u}} &= \frac{1}{Z^\prime} \left( \frac{p^\prime_2}{\omega^\prime - p^\prime_1} b_{u} + b_{d} \right)\ , \\ 
\overline{b^\prime_{d}} &= \frac{1}{Z^\prime} \left( -\frac{p^\prime_2}{\omega^\prime + p^\prime_1} b_{u} + b_{d} \right)\ ,
\end{align}
with $\omega^\prime = \sqrt{{p^\prime_1}^2+{p^\prime_2}^2}$ and $Z^\prime$ a normalized constant. 
Eq (\ref{fineq2}), expressed in terms of $\overline{\Psi}$, then becomes
\begin{equation}
\begin{aligned}
&\partial_s \overline{\Psi} + v(\alpha,\beta)\sigma_z \partial_x \overline{\Psi} = 0\ ,\\
v_2(\alpha,\beta)  =& \frac{1}{3} \left| (c_{2 \alpha -\beta }+2 c_{\beta} \right|..
\label{eq:v2}
\end{aligned}
\end{equation}
This PDE implies ballistic propagation of the quantum walker for fixed $\alpha$ and $\beta$. Similar results are obtained 
when other periodic sequences of finite length are considered, with each result yielding a similar differential equation, albeit with varying functions for $v$, 
as seen in Fig. \ref{fig:vel2}.
As in the previous section, the latter equation can be reformulated in the following covariant form
\begin{equation}
i (\gamma^0 \partial_0 + \gamma^1 \partial_{1}) \overline{\Psi}=0\ ,
\end{equation}
where $\gamma^0 = \sigma_x$, $\gamma^1 = -i \sigma_y$, $\partial_0=\partial_s, \partial_1=\partial_{\tilde{x}}$, while the rescaled coordinate is now $\tilde{x} = x / v_2(\alpha, \beta)$. 
In this rescaled time-space, the velocity is $1$ and the FDTQW-II can again be interpreted as a massless Dirac equation in the $(1+1)$ dimension. 

By taking the continuous limit, we have characterized the propagation behavior of the DTQWs based on both the Fibonacci sequence of the coin operators and the periodic extension of these sequences. Our analysis shows that both of these quantum walks are ballistic and that the continuous limits can reduce to the $(1+1)$-dimensional 
massless Dirac equation. There still remain several unexplored topics, however. For example, we have not examined the continuous limit on the infinite sequence of step operators, as originally defined in Ref.~\cite{Ribeiro2004}. 
The propagation properties of this model were numerically analyzed in \cite[Fig. 3]{Ribeiro2004}.  According to Ref. \cite{Ribeiro2004}, this walk is either ballistic or sub-ballistic depending on the 
initial values of $\alpha$ and $\beta$. Other unknown aspects include the type of transition that occurs when $\tau$ increases towards infinity, and the mechanism by which the behavior of the periodic extension 
changes into that of the infinite sequence.

%%%%%%%%%%%%%%% Acknowledgments   %%%%%%%%%%%%%%%%%%%%%%%%%%%%%%%%%%%%
The authors acknowledge useful discussion with Fabrice Debbasch and Jingbo Wang. 
G.D.M. expresses thanks to the JSPS summer program (SP14203) for financial support, while
L.H. and B.L. thank the Department of Education of the Australian Government for financial support through the AsiaBound Grants Program.
T.W. thanks the JSPS KAKENHI 25400188 and 
IMS Joint Study Program for financial support, while Y.S. thanks the DAIKO Foundation, also for financial support.


\begin{thebibliography}{38}%
\bibitem{Feynman1965} R. P. Feynman and A. R. Hibbs, Quantum Mechanics
and Path Integrals (McGraw-Hill Book Company, 1965).

\bibitem{Aharonov1993} Y. Aharonov, L. Davidovich, and N. Zagury, 
Phys. Rev. A \textbf{48}, 1687 (1993).

\bibitem{Meyer1996}D. A. Meyer, J. Stat. Phys. \textbf{85}, 551 (1996).

\bibitem{Gudder1988}S. P. Gudder, Quantum Probability (Academic Press,1988).

\bibitem{Cardano2014}F. Cardano, F. Massa, H. Qassim, E. Karimi, S. Slussarenko,
D. Paparo, C. de Lisio, F. Sciarrino, E. Santamato,
R. W. Boyd, and L. Marrucci, arXiv:1407.5424

\bibitem{Broome2010}M. A. Broome, A. Fedrizzi, B. P. Lanyon, I. Kassal,
A. Aspuru-Guzik, and A. G. White, Phys. Rev. Lett. \textbf{104}, 153602 (2010).

\bibitem{Kitagawa2012}T. Kitagawa, M. A. Broome, A. Fedrizzi, M. S. Rudner,
E. Berg, I. Kassal, A. Aspuru-Guzik, E. Demler, and
A. G. White, Nat. Commun. \textbf{3}, 882 (2012).

\bibitem{Schreiber2012}A. Schreiber, A. Gabris, P. P. Rohde, K. Laiho, M. Stefanak,
V. Potocek, C. Hamilton, I. Jex, and C. Silberhorn,
Science \textbf{336}, 55 (2012).

\bibitem{Schmitz2009}H. Schmitz, R. Matjeschk, C. Schneider, J. Glueckert,
M. Enderlein, T. Huber, and T. Schaetz, Phys. Rev. Lett. \textbf{103}, 090504 (2009).

\bibitem{Zahringer2010}F. Z\"{a}hringer, G. Kirchmair, R. Gerritsma, E. Solano,
R. Blatt, and C. F. Roos, Phys. Rev. Lett. \textbf{104}, 100503 (2010).

\bibitem{Schreiber2010}A. Schreiber, K. N. Cassemiro, V. Potocek, A. Gabris,
P. J. Mosley, E. Andersson, I. Jex, and C. Silberhorn,
Phys. Rev. Lett. \textbf{104}, 050502 (2010).

\bibitem{Karski2009}M. Karski, L. Forster, J. M. Choi, A. Steffen, W. Alt,
D. Meschede, and A. Widera, Science \textbf{325}, 174 (2009).

\bibitem{Sansoni2012}L. Sansoni, F. Sciarrino, G. Vallone, P. Mataloni,
A. Crespi, R. Ramponi, and R. Osellame, Phys. Rev. Lett. \textbf{108}, 010502 (2012).

\bibitem{Sanders2003}B. C. Sanders, S. D. Bartlett, B. Tregenna, and P. L.
Knight, Phys. Rev. A \textbf{67}, 042305 (2003).

\bibitem{Perets2008}H. B. Perets, Y. Lahini, F. Pozzi, M. Sorel, R. Morandotti,
and Y. Silberberg, Phys. Rev. Lett. \textbf{100},170506 (2008).

\bibitem{Crespi2013}A. Crespi, R. Osellame, R. Ramponi, V. Giovannetti,
R. Fazio, L. Sansoni, F. De Nicola, F. Sciarrino, and
P. Mataloni, Nat. Photonics \textbf{7}, 322 (2013).

\bibitem{Jeong2013}Y. C. Jeong, C. Di Franco, H. T. Lim, M. S. Kim, and
Y. H. Kim, Nat. Commun. \textbf{4}, 2471 (2013).

\bibitem{Fukuhara2013}T. Fukuhara, P. Schaus, M. Endres, S. Hild, M. Cheneau,
I. Bloch, and C. Gross, Nature \textbf{502}, 76 (2013).

\bibitem{Xue2014}P. Xue, H. Qin, B. Tang, and B. C. Sanders, New J. Phys. \textbf{16}, 53009 (2014).

\bibitem{Manouchehri2014} K. Manouchehri and J. B. Wang, Physical Implementation of Quantum Walks (Springer-Verlag, Berlin, 2014).

\bibitem{DiMolfetta2014}G. Di Molfetta, M. Brachet, and F. Debbasch, Physica A \textbf{397}, 157
(2014).

\bibitem{Shikano2010}Y. Shikano, K. Chisaki, E. Segawa, and N. Konno, Phys. Rev. A \textbf{81}, 062129 (2010).

\bibitem{Chandrashekar2008}C. M. Chandrashekar and R. Laflamme, Phys. Rev. A \textbf{78}, 022314 (2008).

\bibitem{Ambainis2007}A. Ambainis, SIAM J. Comput. \textbf{37}, 210 (2007).

\bibitem{Magniez2006}F. Magniez, A. Nayak, J. Roland, and M. Santha, SIAM J. Comput. \textbf{40}, 142 (2006).

\bibitem{Bose2003}S. Bose, Phys. Rev. Lett. \textbf{91}, 207901 (2003).

\bibitem{Aslangul2005}C. Aslangul, J. Phys. A-Math. Gen. \textbf{38}, 1 (2005).

\bibitem{Oka2005}T. Oka, N. Konno, R. Arita, and H. Aoki, Phys. Rev. Lett. \textbf{94}, 100602 (2005).

\bibitem{Bose2007}S. Bose, Contemp. Phys. \textbf{48}, 13 (2007).

\bibitem{Kitagawa2010}T. Kitagawa, M. S. Rudner, E. Berg, and E. Demler,
Phys. Rev. A \textbf{82}, 033429 (2010).

\bibitem{Strauch2007}F. W. Strauch, J. Math. Phys. \textbf{48}, 082102 (2007).

\bibitem{Bracken2007}A. J. Bracken, D. Ellinas, and I. Smyrnakis, Phys. Rev. A \textbf{75}, 022322 (2007).

\bibitem{Sato2010}F. Sato and M. Katori, Phys. Rev. A \textbf{81}, 012314 (2010).

\bibitem{Chandrashekar2010}C. M. Chandrashekar, S. Banerjee, and R. Srikanth,
Phys. Rev. A \textbf{81}, 062340 (2010).

\bibitem{DiMolfetta2011}G. Di Molfetta and F. Debbasch, J. Math. Phys. \textbf{53}, 123302 (2011).

\bibitem{DiMolfetta2013}G. Di Molfetta, M. Brachet, and F. Debbasch, Phys. Rev. A \textbf{88}, 042301 (2013).

\bibitem{Shikano2013}Y. Shikano, J. Comput. Theor. Nanos. \textbf{10}, 1558 (2013).

\bibitem{Ribeiro2004}P. Ribeiro, P. Milman, and R. Mosseri, Phys. Rev. Lett. \textbf{93}, 190503 (2004).

\bibitem{Romanelli2009}A. Romanelli, Physica A \textbf{388}, 3985 (2009).
\end{thebibliography}
\end{document}